\documentclass[proceedings]{JHEP3} 
\PrHEP{ hep2001}

\usepackage{epsfig,multicol}			

\newbox\mybox
\newcommand\fverb{\setbox\mybox=\hbox\bgroup\verb}
\newcommand\fverbdo{\egroup\medskip\noindent\fbox{\unhbox\mybox}\ }
\newcommand\fverbit{\egroup\item[\fbox{\unhbox\mybox}]}
\newcommand{\openone}{\lea{\it%
vev}mode\hbox{\small1\kern-4.2pt\normalsize1}} 
\newcommand{\rr}[4]{#1, {\it #2 \/}{\bf #3} #4}
\newcommand{\Tau}{{\cal T}}
\newcommand{\al}{\alpha}
\newcommand{\alef}{\al_{eff}'}

\newcommand{\gm}{\gamma}
\newcommand{\dl}{\delta}
\newcommand{\xd}{\dot{X}}
\newcommand{\vb}{\bar{v}}

\newcommand{\tr}{\mbox{\rm tr }}

\newcommand{\qb}{\bar{q}}
\renewcommand{\th}{\theta}

\newcommand{\lra}{\longrightarrow}
\newcommand{\sg}{\sigma}

\newcommand{\alp}{\alpha'}
\newcommand{\xpr}{x_\perp}
\newcommand{\f}[2]{\frac{#1}{#2}}
\newcommand{\eq}{\begin{equation}}
\newcommand{\eqx}{\end{equation}}
\newcommand{\eqn}{\begin{eqnarray}}
\newcommand{\eqnx}{\end{eqnarray}}

\newcommand{\DD}{{\cal D}}

\newcommand{\cor}[1]{\left\langle{#1}\right\rangle}
\renewcommand{\AA}{{\cal A}}
\renewcommand{\DD}{{\cal D}}

\newcommand{\ttl}{\f{\tau^2 \th^2}{L^2}}
\newcommand{\Ttl}{\f{T^2 \th^2}{L^2}}

\newcommand{\AmS}{{\protect\the\textfont2A\kern-.1667em\lower.5ex\hbox{M}\
kern-
1
25emS}}

\title{Geometry of Reggeized  amplitudes from AdS/CFT}

\author{\speaker{Robi Peschanski}\\
CEA/DSM/SPhT,Unit\'e de recherche 
associ\'ee 
au CNRS, \\
CE-Saclay, F-91191 Gif-sur-Yvette Cedex,France\\
	E-mail: \email{pesch@spht.saclay.cea.fr}}

\conference{26$^{\rm th}$ Johns Hopkins Workshop}

\abstract{
String theory has long ago  been initiated by the quest for a theoretical 
explanation of   the observed high-energy ``Reggeization'' of  strong 
interaction amplitudes. In terms of quantum field theory, it is  the 
so-called ``soft'' regime, where the coupling constant is expected to be 
large and thus perturbative calculations inadequate. However, since then, 
no convincing derivation of the link between gauge field theory at strong 
coupling and string theory has come out.  This  35-years-old puzzle is 
thus still unsolved.  We discuss how modern tools like the  AdS/CFT 
correspondence give a new insight on the problem by applying it to two-body 
elastic and inelastic scattering amplitudes. We obtain  a geometrical 
interpretation of Reggeization and its relation with confinement in gauge 
theory.}

\begin{document}

\section{Introduction}

\FIGURE{\epsfig{file=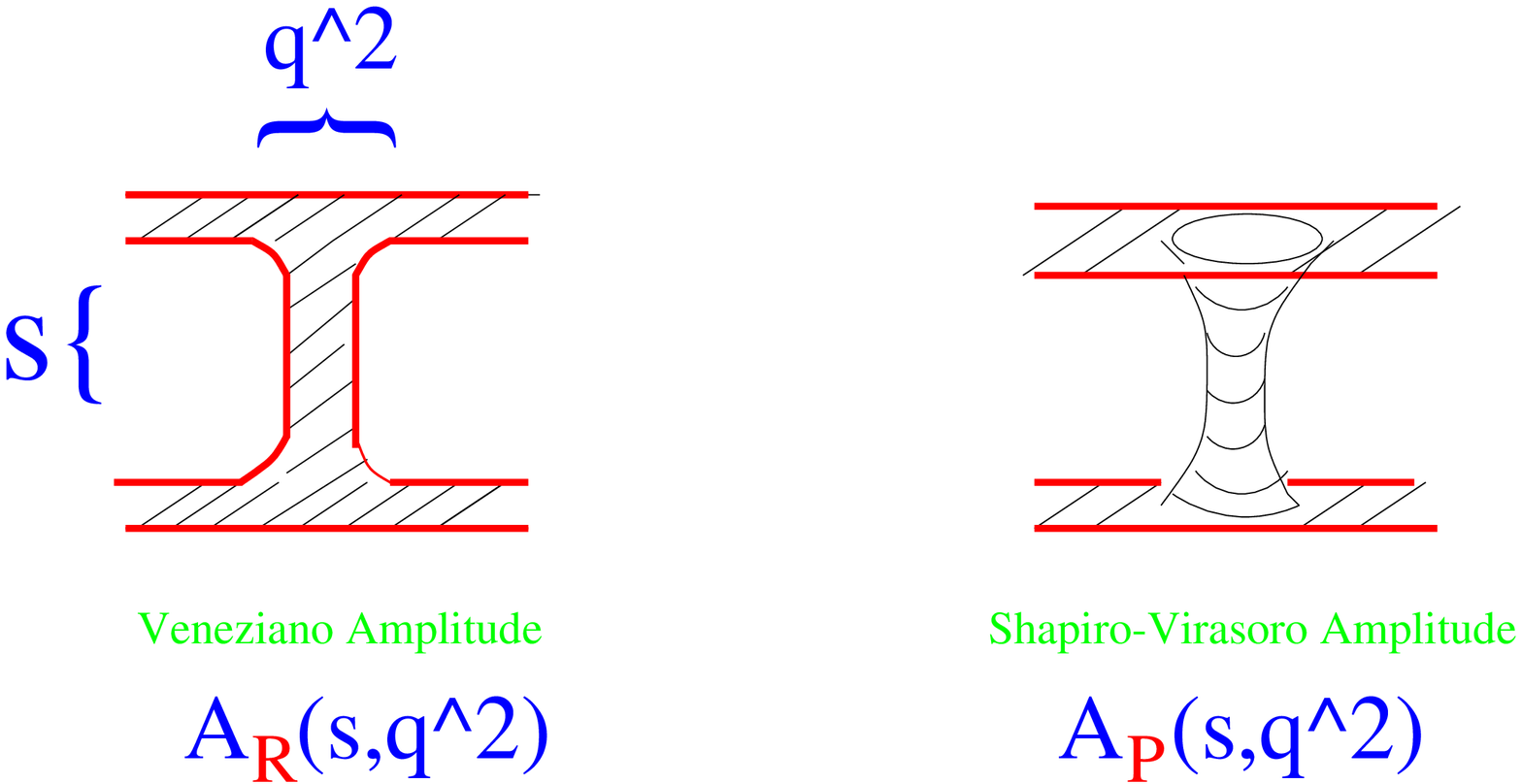,angle=0,width=12cm}%
 \caption{``Duality diagrams'' for two-body inelastic and elastic amplitudes.}%
	\label{1}}

%
It is well-known that string theory started from the proposal of  
scattering amplitudes which may grasp the two major 
structures of soft interaction phenomenology for $2\to 2$ reactions in a 
condensed form: 
resonances and Regge poles. Two  types of amplitudes were proposed 
for four-point amplitudes. The Veneziano amplitude 
corresponds to {\it Reggeon}  exchanges with non-vacuum quantum 
numbers, {\it i.e.} inelastic two-body reaction amplitudes,  and the 
Shapiro-Virasoro 
amplitude  corresponds to {\it 
Pomeron} exchange with vacuum quantum numbers, {\it i.e.} elastic 
amplitudes. These amplitudes have been 
conveniently represented by ``duality diagrams'', see Fig.\ref{1}. In the 
representation of the Veneziano amplitude in terms of quark lines,    
$q\bar q$ intermediate states in the direct $s$-channel correspond to the 
resonances, and $q\bar q$ intermediate states in the exchanged  
$s$-channel to Regge poles, where $s,t$ are the well-known Mandelstam 
variables. The quark lines are to be considered as boundaries of a 
surface 
bearing no quantum numbers, as can be seen for the representation of the 
Shapiro-Virasoro amplitude corresponding to no quantum nmber exchange.

Quite amazingly,  this representation found its justification in terms of 
string theory. The ``duality diagram'' representation of Fig. \ref{1} can be 
mapped into the fusing and splitting of  strings. More precisely, 
the Veneziano and Shapiro-Virasoro amplitudes are topologically related 
to   open and closed tree-level  string respectively. This topological relation 
can be more 
formally expressed in string theory by the invariance of amplitudes with 
respect to deformations of the world sheet spanned by the interacting 
string, and by its conformal invariant properties. After a generalization of 
dual amplitudes has been found for 
multiparticle amplitudes, this raised the hope to find both a 
theoretically 
consistent 
theory of strong interactions and a calculation procedure using the 
perturbative topological expansion of string amplitudes \cite{Frampton}.

However, despite many efforts during years, no widely recognized  
progress has been done in the string theory of strong interaction 
amplitudes. Moreover,  after the discovery of QCD as the gauge field 
theory of quarks and gluons and its validity for  a quantitative 
description of many ``hard'' scattering processes, there remained little 
hope that a string theory of strong interactions could take place. Indeed, 
even 
before the discovery of QCD,   major theoretical obstacles  to the 
formulation 
of 
a consistent string theory of strong 
intractions were being  
raised. Let us give a non exhaustive list of problems.

 The conformal anomaly of string theories in Minkowski $D$-dimensional 
space leads to the limitation  $D=26,10$ for respectively bosonic and 
super strings in flat space. More generally, the requirement of conformal 
and diffeomorphism invariance imposes stringent  constraints on the space 
in which the string moves. 

Zero mass gauge and  gravitational fields appear in the  string spectra 
of 
 asymptotic states. Consistent string theory, when considered  in flat 
target space, contains (in general) gauge group and gravitational fields 
and degrees of freedom. They are thus more suitable  for 
a 
stringy approach of grand unified theories, than for strong interactions 
and the confinement problem, characterized by the absence of asymptotic 
zero mass states  in the theory.

To these  difficulties, new ones  have been added  after  the discovery 
of 
QCD. Let us list three among the main questions (at least those concerning 
the 
scattering of two particles): 

\begin{itemize}
\item {\it Can we find a consistent  picture of the Reggeization of 
high-energy amplitudes when QCD enters its strong coupling regime?}. Even 
knowing the QCD lagrangian, it has not been possible to derive from it 
high-energy amplitudes in the soft interaction case. Lattice calculations 
have been useful for investigations at low energy but are seemingly 
hopeless  in the kinematical domain of high energies. 
\item {\it Where  are ``hard'' interactions recovered in a 
 string 
theory framework?} String world sheets are suitable objects for the 
description of ``soft'' phenomena due to their extended structure in 
space 
and time. It is more intricate to show off in string theory the ``hard'' 
structure visible in short-time interactions. 
\item {\it Can we elaborate a suitable string theory which 
could  coherently describe the  properties  of gauge fields?}. The 
degrees 
of freedom and the symmetries of a gauge field theory are much different 
from those commonly found for string theories. the matching of these two 
require  non trivial constraints as recognized in particular in 
Ref.\cite{polyakov}.
\end{itemize}

To the first of these three  questions, the recently proposed duality 
corespondence
between certain string and certain gauge field theories gives new and 
reliable 
answers. It seems that the second one can find some interesting clues 
also in 
the same framework
\cite{17}. we will focus here upon  the third one, namely  the 
understanding 
of  Regge two-body amplitudes in gauge field theories at strong coupling 
in 
terms 
of the AdS/CFT duality. This question  has been the subject of an 
approach 
\cite{us1,us2,us4,us3} which I will now develop.

The plan of the present review is the following: in section {\bf 2}, we 
will 
give a brief account of the  AdS/CFT dual correspondence, focusing on  
aspects 
relevant for our study. In section {\bf 3}, we will explain the formalism 
using the classical approximation and minimal surfaces  for the determination of 
the AdS duals of  two-body scattering amplitudes. We successively apply it to 
quark (anti)quark, dipole-dipole elastic scattering and finally inelastic   
dipole-dipole scattering. Next in {\bf 4}, we will develop a semi-classical 
approximation, improving the previous method by computing the fluctuation factor 
around the minimal surface solution. A final section {\bf 5} is devoted to a 
summary and conclusion about the geometrical nature of   Reggeization in the 
confining AdS dual backgrounds.

\section{String/Gauge fields Duality}

The AdS/CFT correspondence \cite{11}  has many interesting formal and 
physical facets. Concerning 
the 
aspects which are of interest for our problem, it allows one to find 
relations 
between gauge field theories at strong coupling and string gravity at 
weak 
coupling  in the limit of large number of colours ($N_c\!\to\! \infty$). It can 
be examined  quite precisely in  the 
AdS$_5$/CFT$_4$ 
case which conformal field theory  corresponds to $SU(N)$ gauge 
theory 
with ${\cal N} \!=\!4$ supersymmetries. 

Some existing extensions to other gauge theories with  broken conformal 
symmetry
and less or no supersymmetries will be valuable for our approach, since 
they 
lead to confining gauge theories  which 
are 
more similar to  QCD\footnote{Note  that the 
appropriate 
string gravity dual of QCD has not yet been identified, and thus we are 
forced 
to restrict for the moment our  use of AdS/CFT correspondence to 
features 
which 
are 
expected to be a general feature of  confining theories duals, see a 
discussion 
further on in this section.}. Indeed, one important question is  to examine to 
what 
extent confinement plays a r\^ole in the Reggeization of 
amplitudes. Our aim is thus to investigate the possible 
realization and origin of 
Reggeization of two-body amplitudes in such theories and what are the 
difference 
appearing with the original AdS$_5$/CFT$_4$ case.

\FIGURE[ht]{\epsfig{file=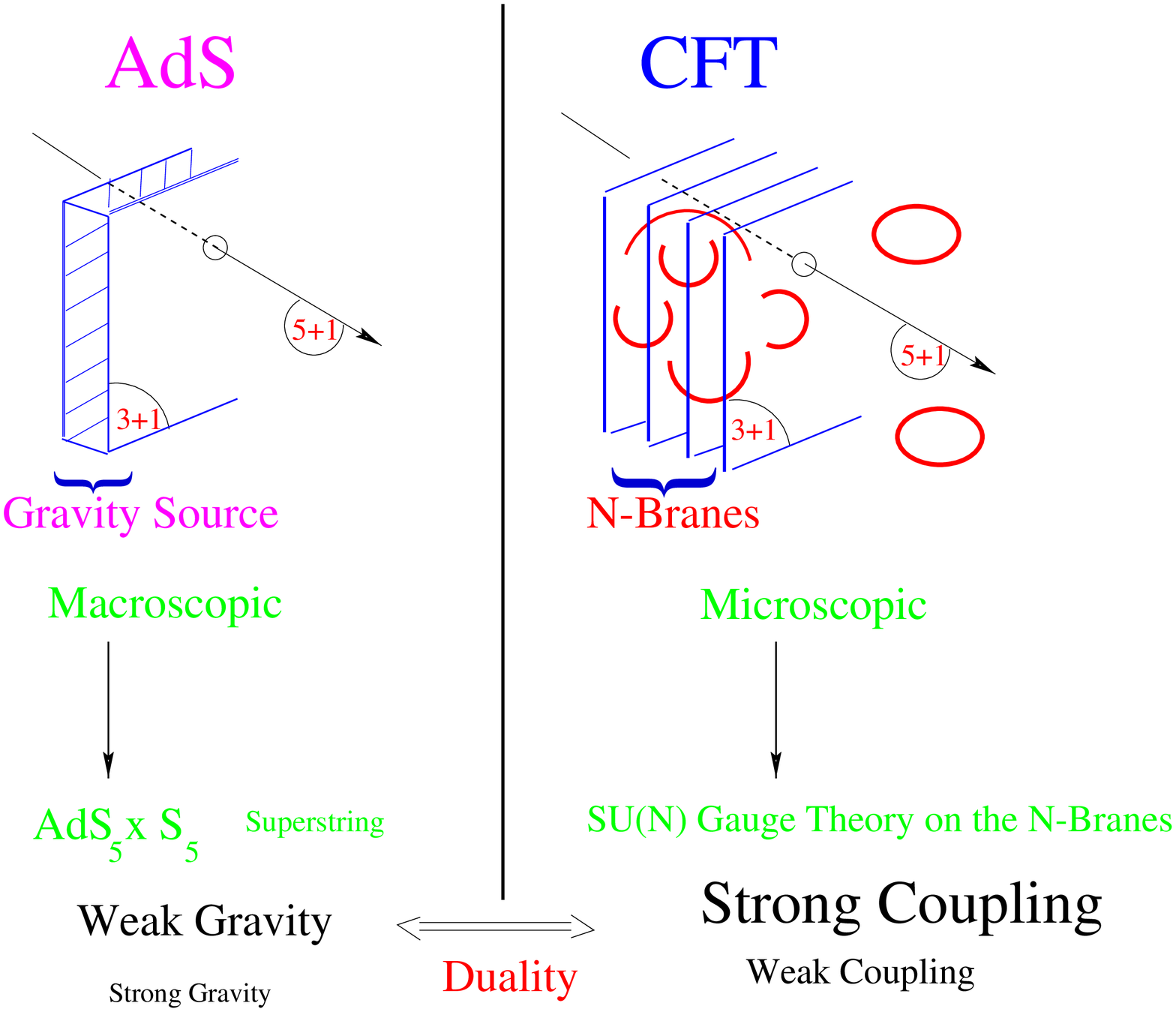,angle=0,width=15cm}%
 \caption{AdS$_5$/CFT$_4$ duality correspondence.}%
	\label{2}}

Let us  recall the canonical derivation leading to  the AdS$_5$ 
background \cite{review}, 
see 
Fig.\ref{2}. One starts from  the (super)gravity classical 
solution of a system of $N\ D_3$-branes in a $10\!-\!D$ space of the (type 
IIB) 
superstrings. 
The metrics solution of the (super)Einstein equations read
\eq
\label{super}
ds^2=f^{\!-\!1/2} (\!-\!dt^2\!+\!\sum_{1\!-\!3}dx_i^2) 
\!+\!f^{1/2}(dr^2\!+\!r^2d\Omega_5) 
\ ,
\eqx
where the first four coordinates are on the brane and $r$ corresponds to 
the  
coordinate along the normal to the branes. In formula (\ref{super}), one defines
\eq
f=1+\frac {R^4}{r^4}\ ;\ \ \ \ \ R=4\pi g^2_{YM}\alpha'^2 {N} \ ,
\label{R}
\eqx
where $g^2_{YM}{N}$ is the `t Hooft-Yang-Mills coupling and $\alpha'$ the 
string 
tension. 

One considers the   limiting behaviour considered by 
Maldacena, 
where 
one zooms on the  neighbourhood of the branes  while in the same time going to 
the limit of 
weak 
string 
slope $\alpha'.$ The near-by space-time is thus distorted due to the 
(super) 
gravitational field of the branes. One goes to the limit where 
\eq
\ R\ fixed\ ; \frac {\alpha'(\to 0)} {r (\to 0)}\to z\ fixed\ .
\eqx
This, from the second equation of (\ref{R}) obviously implies 
\eq
{\alpha'\to 0}\ , \ g^2_{YM} 
{N}\sim \f{1}{\alpha'^2}\to \infty \ ,
\eqx
{\it i.e.} both a weak coupling  limit for the string theory and   a strong 
coupling limit for the dual gauge field theory.
By reorganizing the two parts of the metrics one obtains
\eq
ds^2={ \frac 1{z^2} (-dt^2+\sum_{1-3}dx_i^2+ dz^2)} +  {R^2 
d\Omega_5}\ ,
\label{AdS}
\eqx
which corresponds to the 
{AdS$_5$} $\times \  {S_5}$ background structure,  ${S_5}$ being  
the 5-sphere. More detailed analysis shows that the isometry group of 
the  
5-sphere is the geometrical dual of the ${\cal N}\! =\!4$ 
supersymmetries. More intricate is the quantum number dual to $N_c,$ the number 
of 
colours, which is the invariant charge carried by the Ramond-Ramond form 
field.

In the case of confining backgrounds, an intrinsic scale breaks conformal 
invariance and is brought in the dual theory  through {\it e.g.} a 
geometrical constraint. For instance in~\cite{wi98} a proposal was made 
that a 
confining gauge theory is dual to string theory in an $AdS_{BH}$ black
hole (BH) background 

\eq
\label{e.bhmetric}
ds^2_{BH}=\f{16}{9}\f{1}{f(z)}\f{dz^2}{z^2} + \f{\eta_{\mu\nu}dx^\mu
dx^\nu}{z^2} + \ldots 
\eqx
where  $f(z)=z^{2/3}(1-(z/R_0)^4)$ and $R_0$ is the position of 
the horizon.
We will use this 
background\footnote{Although it was later found that the $S^1$ KK states do not 
strictly
decouple ~\cite{de99}.} 
to
study the interplay between the confining nature of gauge theory and
its reggeization properties. Actually the qualitative 
arguments and
approximations should be generic for most confining backgrounds, as 
already 
discussed in Ref.~\cite{so99}.
For instance, 
other geometries for (supersymmetric)
confining theories \cite{ks,carlo} have been discussed in this respect. 
They have the property that for
small $z$, i.e. close to the boundary, the geometry looks like 
$AdS_5\times
S^5$ (in \cite{carlo} up to logarithmic corrections related to asymptotic
freedom) giving a coulombic $q\qb$ potential. For large $z$ the
geometry is effectively flat. In all cases there is a scale, similar
to $R_0$ above, which marks a transition between the small $z$ and
large $z$ regimes.

In order to illustrate the way  one formulates the AdS/CFT 
correspondence in a context similar to QCD, 
let us consider the example of  the vacuum expectation value ({\it vev}) 
of 
Wilson lines in a  configuration parallel
 to the time direction of the branes. This configuration
allows 
a   determination of the 
potential between colour charges \cite{12}. The r\^ole of colour charges
in the fundamental representation is played by open string states 
elongated 
between a  stack of $N_c$ $D_3$ branes on one side and one $D_3$ brane 
near the 
boundary of AdS space ({\it cf.} \cite{review}).

\FIGURE[hb]{\epsfig{file=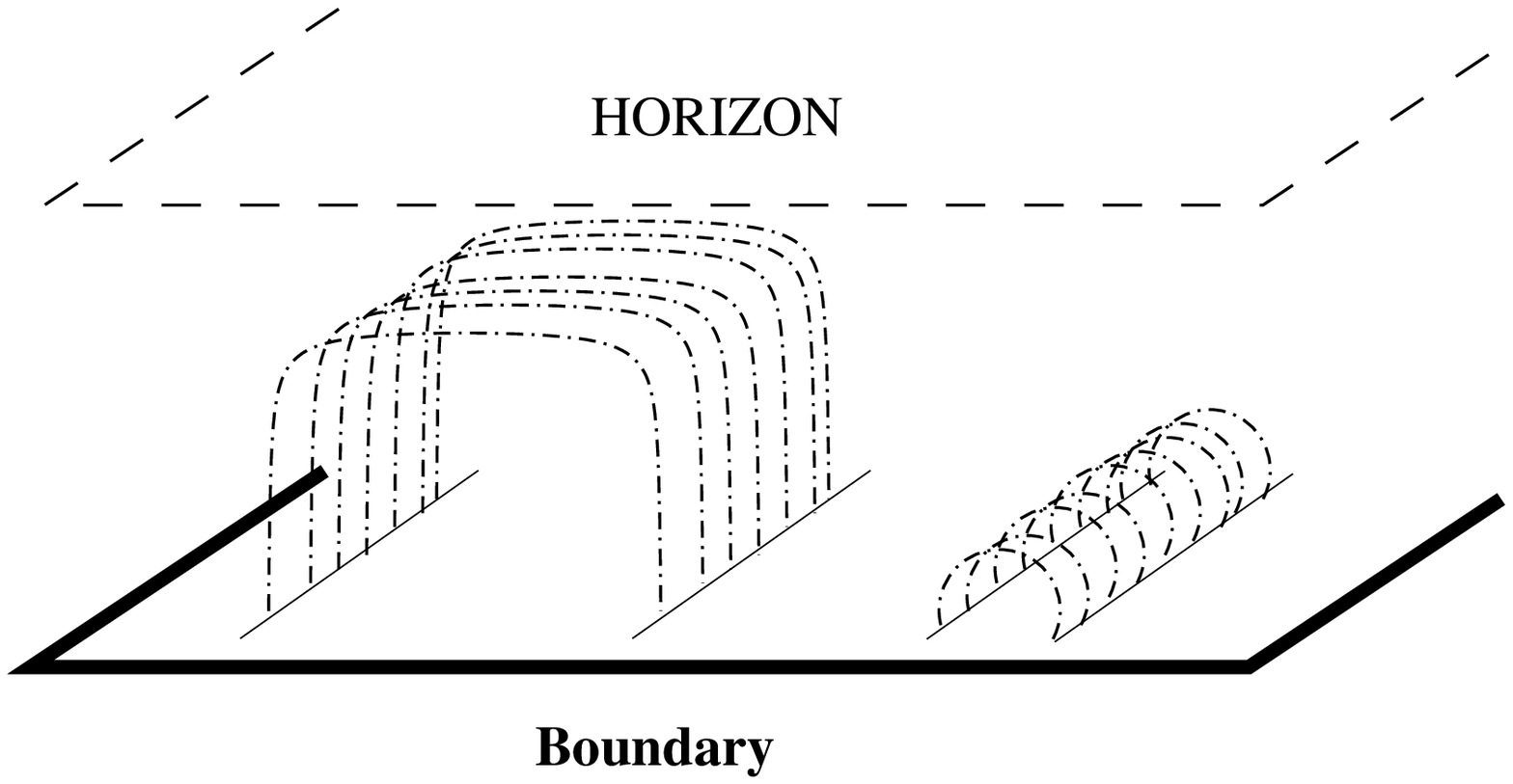,angle=0,width=12cm}%
 \caption{Exemple of minimal surfaces with Wilson line boundaries.}%
	\label{3}}

One writes
\eq
\langle e^{iP\int_C\vec A\cdot\vec dl}\rangle 
\!=\!\!\int_{\Sigma}\!e^{- 
\frac{Area({\Sigma})}{\alpha'}}\!\!\approx e^{-
\frac  {Area_{min}}{\alpha'}}\times 
Fluct.\ ,
\eqx
where $C$ is the Wilson line contour near the $D_3$ branes  and $\Sigma$ 
the 
surface in $AdS$-space with $C$ as the boundary, see Fig.\ref{3}. The 
minimal 
area 
approximation is  the {\it vev} evaluation in the classical $\alpha'\to 0$ 
limit. The factor denoted $Fluct.$ refers to the  fluctuation 
determinant  
around the minimal surface, corresponding to the first one-loop (in 
$\sigma$-model sense) quantum correction. It gives  a calculable semi-classical 
correction.

  In  Fig.\ref{3}, we have 
sketched  the form of  minimal surface solutions for the  
``confining'' $AdS_{BH}$
case,   (see 
above (\ref{e.bhmetric})). For large separation of Wilson lines, the minimal 
surface ``feels'' the horizon and is consequently curved. At smaller separation, 
the solution becomes again similar to the conformal case, since the horizon 
cut-off does not play a big r\^ole.

The {\it vev} results in the classical approximation can be summed up as 
follows:
\eqn
AdS_5: \langle Wilson\ Lines\rangle&=&e^{TV(L)}\sim 
e^{\#_1T/L}\nonumber \\ AdS_{BH}: \langle Wilson\ 
Lines\rangle&=&e^{TV(L)}\sim e^{\#_2TL/R_0^2}\ , \nonumber 
\eqnx 
where, the potential behaviour is as expected for respectively conformal 
(perimeter law) and confining (area law) cases.   Note that  there is an   
interesting  information  stored in the coupling dependent numbers 
here denoted by  $\#_{1,2}$ . Note also that, even in the case of a 
confining 
geometry with an horizon at $R_0,$ Wilson lines separated by a distance 
$L<<R_0 
$ do not give rise to minimal surfaces  sensitive to the horizon (see 
Fig.3), 
and thus give rise to classical solutions similar to the non-confining case.

The important r\^ole of  fluctuation corrections and the way of computing 
it 
in some non-trivial cases will be discussed further on.

\section{Supergravity Duals of Scattering Amplitudes}

Using the AdS/CFT correspondence, we find that two-body high energy amplitudes 
in gauge field 
theories can 
be 
related to specific configurations of 
minimal surfaces\footnote{A different approach has been independently 
proposed  in Ref.\cite{13}.}. 

Indeed, at high energy, fast moving colour sources 
propagate 
along
linear trajectories in coordinate space thanks to the eikonale 
approximation. An analytic 
continuation 
from Minkowski to Euclidean ${\cal R}^4$ space allows one to find a  
geometrical interpretation in terms of a well-defined minimal 
surface 
problem. Let us 
consider for illustration different applications. 

\subsection{Quark-quark elastic scattering}

\FIGURE[ht]{\epsfig{file=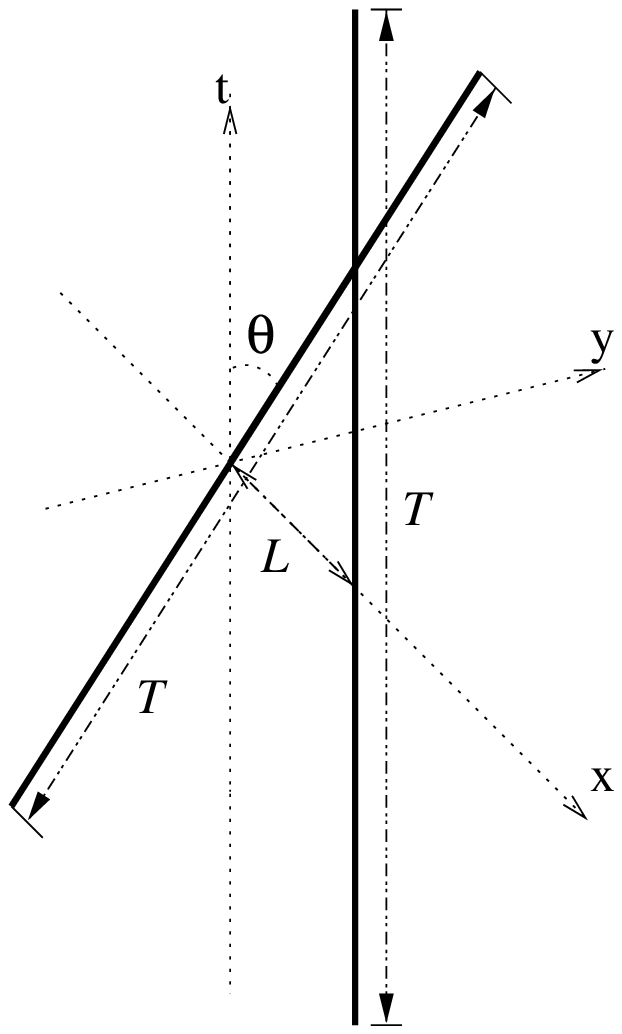,angle=0,width=5cm}%
 \caption{Wilson lines for  quark-quark  elastic scattering in 
${\cal 
R}^4.$}%
	\label{4}}

In a  framework suitable for performing the AdS/CFT correspondence, quarks 
(resp. antiquarks) can be represented by  colour sources in the fundamental  
(resp. 
anti fundamental)
reps. of SU(N). In the brane world, they are obtained ({\it e.g.} 
see 
\cite{review}) by considering systems of $N\!+\!1\ D_3$ branes of which one of 
the 
brane is removed to a distance from the remaining stack of $N\ D_3$ 
branes. This 
distance is large (or equivalently $z$ small) in order to satisfy a static 
approximation for the quarks considered as ultra-massive.

In the corresponding  gauge field theoretical framework, it is knowen 
since a 
long time \cite{14} that the high-energy elastic quark-(anti)quark
amplitude can be formulated as follows
\eqn
A(s,q^2)=2is\int d\vec l\ e^{i\vec q\cdot \vec l}\ \langle W_1 
W_2\rangle 
_{L=|\vec l|}^{\chi=\log s/m^2}\nonumber\\
 = 2is\int d\vec l\ e^{i\vec q\cdot \vec 
l-
\frac  {1}{\alpha'}Area_{min}(\vec l)}\ ,
\label{e.anal}\eqnx
where $\vec l$ is the impact parameter vector between the two 
trajectories, 
conjugated to the momentum transfer $\vec q,$ and $\chi=\log s/m^2$ the 
total 
rapidity 
interval. Performing an analytic 
continuation from Minkovskian to  Euclidean space \cite{meggio}:
\eq
\chi \to i\theta\ \ \ ;\ \ t_{Mink}\to -it_{Eucl}\ ,
\label{anal}
\eqx
the Wilson line {\it vev} can be expressed as a minimal surface problem 
whose 
boundaries are two straight lines in a 3-dimensional coordinate space, 
placed at 
an impact parameter distance $L$ and rotated one with respect to the  other by  
an 
angle 
$\theta,$ 
see 
Fig.4.  In flat space, with the same boundary conditions,  the 
minimal surface is 
the 
{\it helicoid}. One thus  realizes that the problem can be formulated as a 
minimal surface problem whose  mathematically well-defined   solution   is 
a 
{\it generalized  helicoidal} manifold embedded in curved background 
spaces, 
such as Euclidean AdS Spaces. Unfortunately, this problem is rather 
difficult 
to 
solve analytically, even in flat space.  It is known as the Plateau 
problem, 
namely the  determination of minimal surfaces for given boundary conditions 
(see 
for instance \cite{minimal}).

Thus, the interest of considering quark-quark scattering relies on the 
simple 
definition of the  minimal surface geometry in the conditions of a 
confined 
$AdS_{BH}$ metrics (\ref{e.bhmetric}). Indeed, in the 
configuration 
of Wilson lines of Fig.\ref{3} in the context  of a confining theory, the  
AdS$_{BH}$ metrics is characterized by a singularity at $z=0$ which 
implies a 
rapid growth in the  $z$ direction 
towards the D$_3$ branes, then stopped near the horizon at $z_0.$ Thus, 
to a good approximation, and for large enough
impact 
parameter (compared to the horizon distance), the main contribution to 
the 
minimal area is from the  metrics in the bulk near $z_0$ which is nearly
flat. Hence, near $z_0,$ the relevant minimal area can be  drawn on a
classical helicoid which can be parametrized:
\eqn
t &=& \tau \cos {\theta\sg}/{L}\nonumber\\ 
y &=& \tau \sin {\theta\sg}/{L}\nonumber\\
x &=& \sg\nonumber\\  
z &\sim& z_0\ .
\label{helico}
\eqnx

However, the practical calculation \cite{us1} of the amplitude is  
complicated 
by the necessity of introducing a cut-off in the $T$-direction (see 
Fig.4). This 
is physically expected since the area spanned by the helicoid in the 
confining 
geometry goes to infinity, corresponding to the spreading of the 
color 
field between the quarks, the confining forces increasing till the string 
breaks 
for the production of particles, not described in the present scheme. It 
is the 
expected  counterpart, in QCD, of the infinite phase of electron-electron 
scattering in QED. Let us sketch the calculations of  \cite{us1}.

	 The truncated 
helicoid solution is parametrized by (\ref{helico})with $\tau=-T\ldots T,$  
$\sg=0 \ldots L$ and $\th$ is the total
twisting angle.  
Its area is given by the formula
\eq
\label{e.ahelic}
Area =\int_{0}^{L}d\sigma \int_{-T}^{T}d\tau \sqrt{1+\ttl}=
LT\sqrt{1+\Ttl } +\f{L^2}{2\th} \log
\left\{\f{\sqrt{1+\Ttl}+\th \f{T}{L}}{\sqrt{1+\Ttl}-\th \f{T}{L}}\right\} 
\ .
\eqx
Through the analytical continuation (\ref{anal}), one would naively obtain 
a 
pure $T$-dependent phase factor going to $\infty$ when removing the 
cut-off.
However the
analytic structure of the euclidean area (\ref{e.ahelic}) involves 
cuts in the complex $T$, $\th$ planes and thus leads to an ambiguity
coming from the branch cut of the 
logarithm. In fact when performing the analytical continuation we have
to specify the Riemann sheet of the logarithm (i.e. $log \to log+2\pi
i n$). 

This leads \cite{us1} to a $T$-independent real 
term which, inserted in formula (\ref{e.anal}),  gives rise to a reggeized 
amplitude. Doing this,  the 
removal we make of the (infinite as $T\to \infty$)
 phase could be considered as 
a 
natural infrared regularisation. One finally obtains
\eq
\label{e.ai}
A_{\cal P}(s,q^2)= 2is\int d\vec l\ e^{i\vec q\cdot \vec l-\left\{ n 
\f{\sqrt{2g_{YM}^2N}}{\chi} \f{L^2}{2R_0^2}\right\}}  \propto
s^{1-q^2\f{R_0^2}{n\sqrt{8g_{YM}^2N}}}\ .
\eqx
which represents a ($n$-dependent) set of Reggeized elastic 
amplitudes, with linear Regge trajectories characterized by a Regge 
intercept 1 
and Regge slopes given by $\f{R_0^2}{n\sqrt{8g_{YM}^2N}}$. In this 
framework the 
removal of the (infinite as $T\to \infty$) phase could be considered as a 
natural infrared regularisation. We will see now how this assumption, 
related to 
the consideration of unphysical asymptotic quark states,  can be relaxed 
without 
affecting the Reggeization property, when considering  scattering between 
colorless dipoles
\vspace{2cm}.

\begin{figure}[ht]
\begin{center}
\includegraphics[width=16pc]{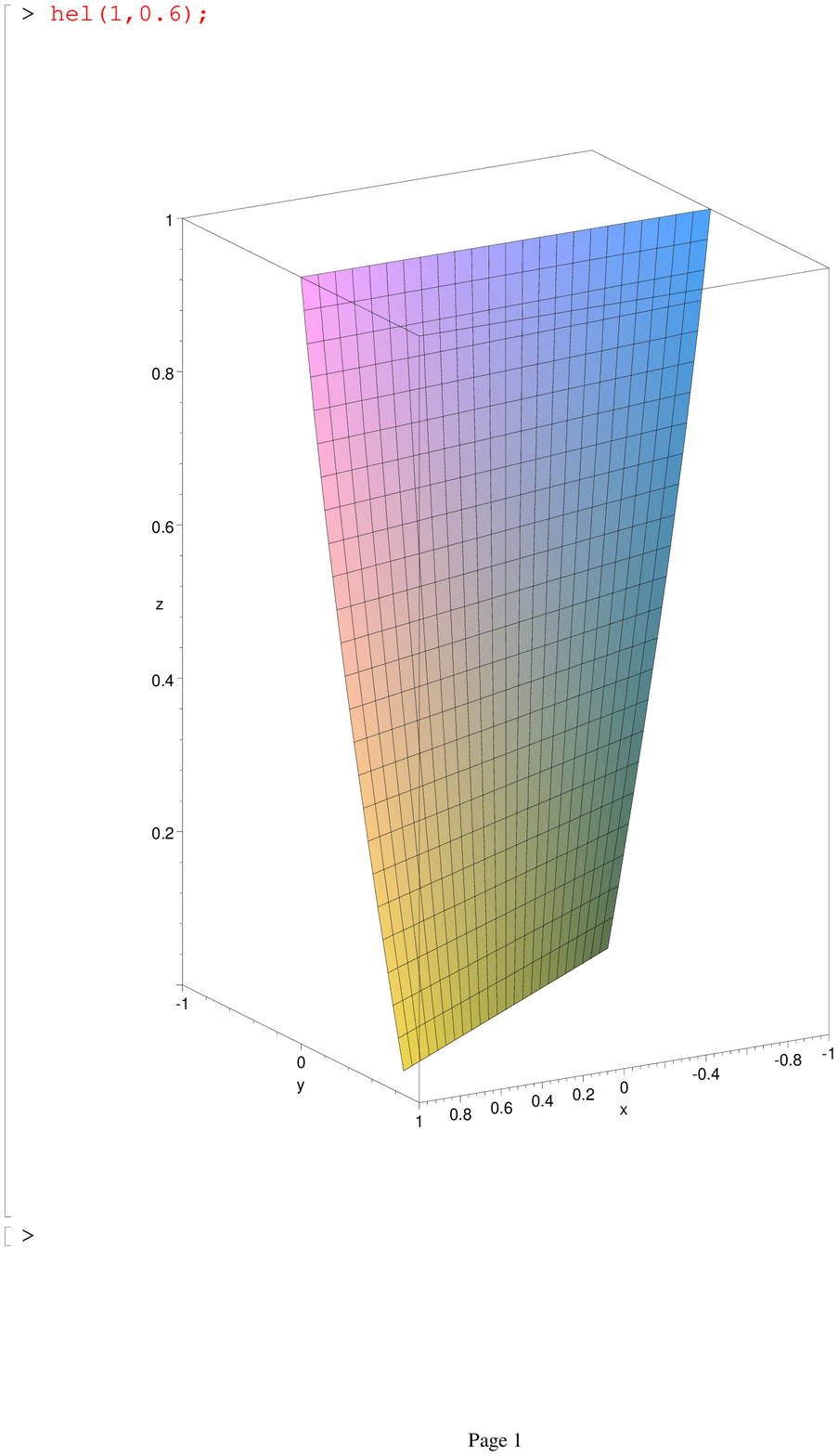}
\includegraphics[width=18pc]{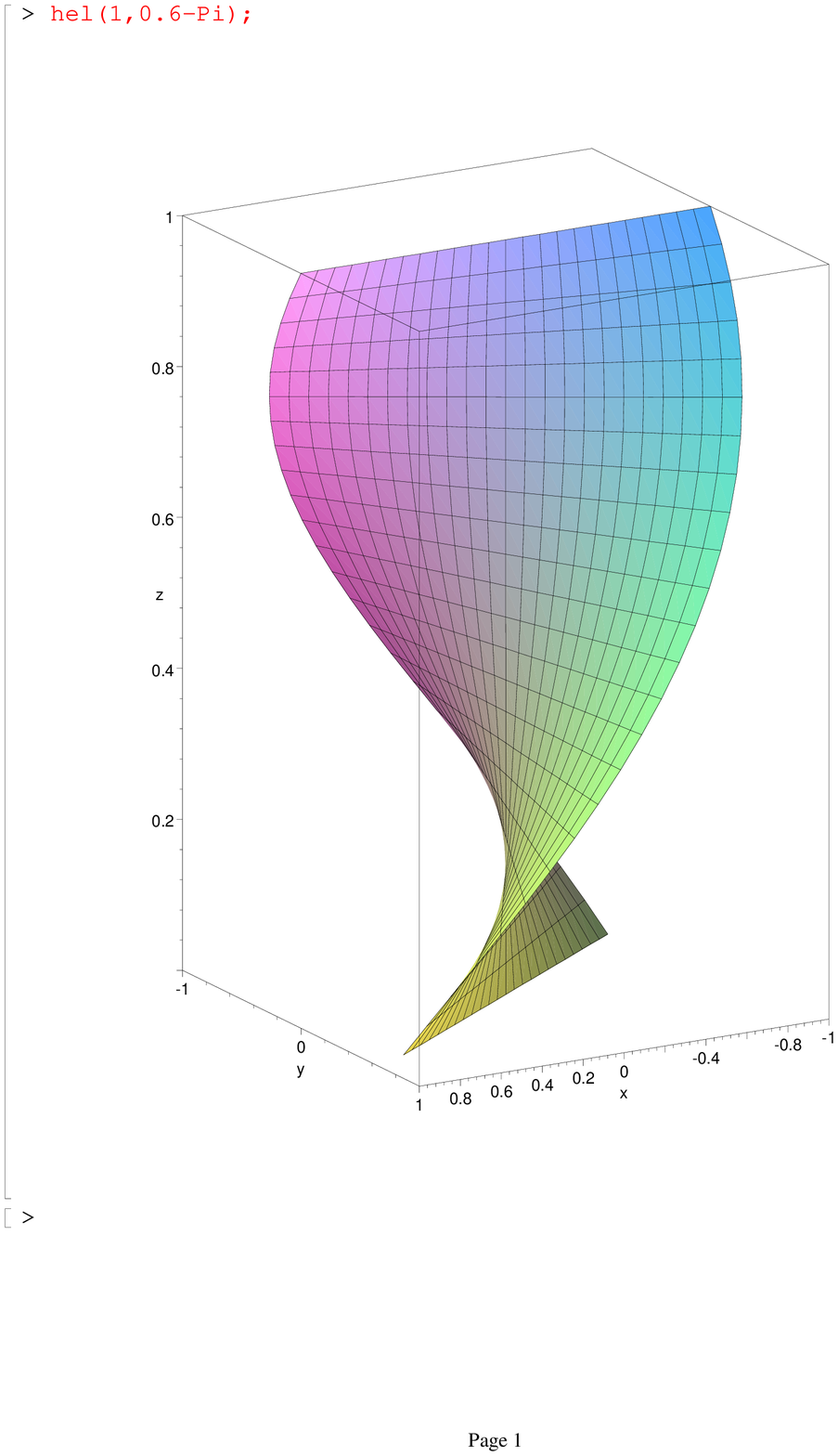}
\vspace{2cm}
\end{center}
\label{45}
\end{figure}
\vspace{7cm}.
{\bf Figure 4bis}: Helicoids describing quark-quark and quark-antiquark 
scattering.
\vspace{2cm}

It is interesting to note that the  realization of Reggeization provided by the 
helicoid geometry through analytic continuation also gives a natural 
interpretation of the ``signature'' factor, {\it i.e.} the phase factor 
distinguishing quark quark scattering and quark antiquark 
scattering in Regge amplitudes. Indeed,  quark quark scattering and quark 
antiquark 
scattering are related through  twisted helicoidal configurations in the bulk 
coordinate 
space. For  a given  helicoid configuration representing  quark quark 
scattering,  twisting one of the  a quark lines will give rise to the  helicoid 
representing quark antiquark 
scattering with the same kinematics, see Fig.4bis.  Hence, through analytic 
continuation, one 
finds
\eq
\th \to \th + \f{\pi}2\  \Rightarrow \ s \to s\ e^{-i\pi}\ ,
\label{signature}
\eqx
which,  once inserted in formula (\ref{e.ai}), gives rise precisely  to the 
Regge phase signature factor.

In the non-confining    $AdS_5 \times S_5$ case, one would need to identify
a 
{\it generalized  helicoidal} manifold embedded with the metrics 
(\ref{AdS}), {\i.e.} the minimal surface with straight line boundaries with an 
angle. 
This is a well-defined mathematical problem, which is yet not solved. 
With 
some crude approximation however, looking for a variational solution with 
$z_0 
\to z(\sigma,\tau)$ in the parametrization (\ref{helico}), one can obtain 
\cite 
{us2} a solution at large $\log s \ :$
\eq
\label{e.lcft}
A(s,q^2) \propto \left(\f{L}{\log s}\right)^{n\f{F(\pi/2)}{\pi} \ 
\f{\sqrt{2g^2_{YM}N}}{2\pi}} 
\  s^{n\f{ 2\pi^4}{\Gamma(1/4)^4} \  \f{\sqrt{2g^2_{YM}N}}{2\pi}} \ ,
\eqx
where $F(\Omega)$ is the ``cusp anomalous dimension'' calculated in \cite{neww}
($F(\pi/2)\sim 0.3 \pi$). It is interesting to note that in this case, 
there is 
no Reggeization, at least with a non-zero Regge slope. Formula 
{\ref{e.lcft}} 
appears as an extension  of the weak coupling result \cite{bfkl} including 
 a 
screening effect on the coupling ($g^2_{YM}N \to \sqrt{g^2_{YM}N}$) when 
it 
becomes strong, as for  the potential in the conformal case \cite{12}.
\subsection{Dipole-dipole  elastic scattering}
\label{dub}

\FIGURE{\epsfig{file=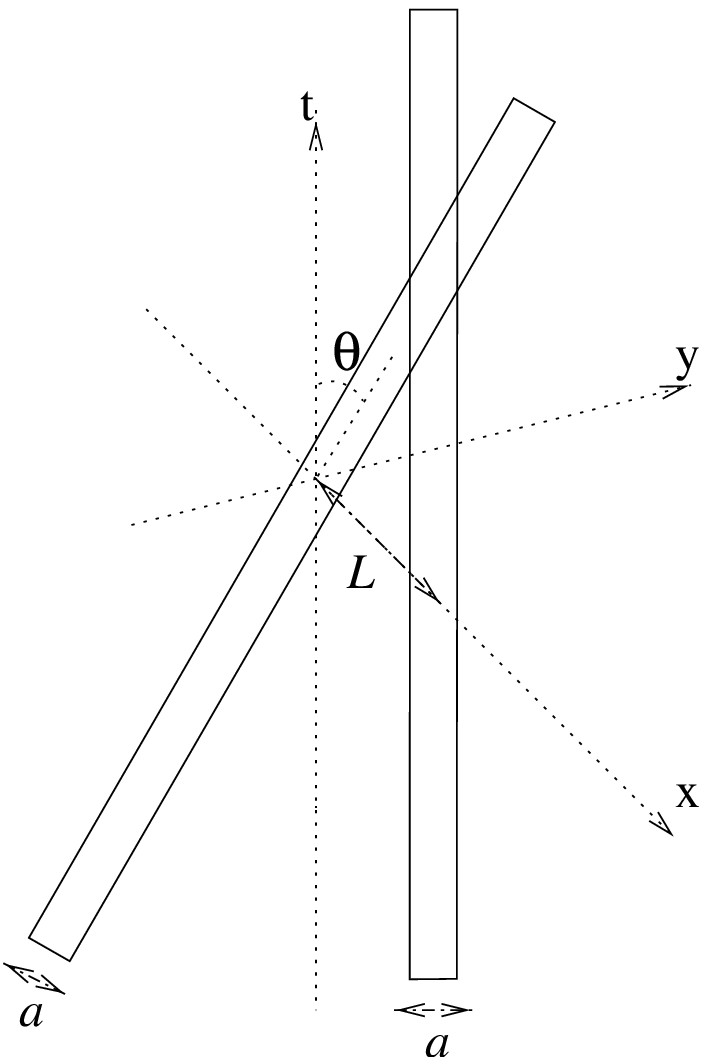,angle=0,width=5cm}%
 \caption{Wilson contours for  Dipole-dipole  elastic scattering.}%
	\label{fig}}

Elastic  scattering of colourless states is expected to be cured from the 
infra-red divergent phase factor encountered in quark-(anti)quark 
scattering. In 
this respect, it is interesting to 
consider the elastic interaction of two very massive 
QCD dipoles. Thanks to their high mass (or equivalently small size $a$), one 
can 
neglect the fluctuations around their classical trajectory and thus  their  
propagation 
in 
coordinate space in the eikonale approximation can be 
represented by elongated Wilson loops near both right and left moving 
light-cone 
directions. More precisely \cite{us1}, one has to 
compute a Wilson loop correlator
 in the configuration displayed in 
Fig.\ref{fig}.
\eq
\label{e.ampinit}
A(s,q^2)=-2is \int d^2\xpr e^{iq\xpr} 
\cor{\f{W_1W_2}{\cor{W_1}\cor{W_2}}-1}
\eqx
where the Wilson loops follow classical straight lines for 
quark(antiquark)
trajectories: $W_1\lra x_1^\mu=p_1^\mu\tau\,(+a^\mu)$ and $W_2\lra
x_2^\mu=\xpr^\mu+p_2^\mu\tau\, (+a^\mu)$ and close at infinite times. The 
normalization ${\cor{W_1}\cor{W_2}}$ of the correlator ensures that the 
amplitude vanishes when 
the 
Wilson loops get decorrelated at large distances.
Let us consider the solution in the confining background (\ref{e.bhmetric}),
approximated by a flat metrics near the horizon. For this setup we have to 
calculate the correlation function of two
Wilson loops 
elongated along the ``time'' 
direction  and have a large but arbitrary temporal length $T$
(the exact analogue for Wilson loops of $T$ considered in
the previous section). However, the cut-off
dependence on  $T$ is  removed and thus  
 the related IR divergence which was present for the $q(\bar q)-q$ scattering 
case. Indeed, for large positive and negative times the minimal 
surface will be well
approximated by two seperate copies of the standard minimal surfaces for
each loop separately. When we come to the interaction region, and for $L$
sufficiently small, one can lower the area by forming a ``tube'' joining
the two worldsheets.
Since we want to calculate the normalized correlator $\cor{W_1
W_2}/\cor{W_1}\cor{W_2}$, the contributions of the regions outside the
tube will cancel out (in a first approximation neglecting deformations
near the tube). Therefore we have just to find the area
of the tube, and subtract from it the area of the two independent
worldsheets. It is at this stage that we see that the result does not
depend on the maximal length of the Wilson loops $T$, and hence is IR
finite. The whole contribution to the amplitude will just come from
the area of the tube.

Our calculation scheme proposed in \cite{us2} goes as follows. Since one 
does 
not know the  explicit minimal surface for these boundary
conditions, let us perform a variational approximation. Namely we 
consider a family of surfaces forming the tube, parameterized by 
$T_{tube}$,
which has the interpretation of an ``effective'' time of
interaction. Then we make a saddle point minimization of the area
as a function of this parameter.

Suppose that the tube linking the two Wilson lines is formed in the
region of the time parameter 
 $t\in$ $(-T_{tube},T_{tube}).$ In our
approximation its two
``sides'' are formed by a duplication of the helicoid solution . The front and 
back 
will
be each approximated by strips of area $aL \sqrt{1+\f{T^2_{tube}
\th^2}{L^2}}$ (we assume $a,L \geq R_0$).

The total area corresponding to the two Wilson loops is then given by 
\eq
\label{e.tube}
Area(T_{tube}) = 2L\int_{-T_{tube}}^{T_{tube}}d\tau \sqrt{1+
\ttl}  +2aL \sqrt{1+\f{T^2_{tube} \th^2}{L^2}} -4a
\cdot T_{tube} \ ,
\eqx
where $-2 a T_{tube}$ is the contribution of each individual Wilson loop
to the normalization $1/\cor{W_1}\cor{W_2}$ of the Wilson loop
correlation function. 

Analytically continuing the area formula (\ref{e.tube}) to the
Minkowskian case and using a convenient change of variables, the
Minkowskian area can be put in the following simple form
\eq
\label{e.tubeminkow}
Area(T_{tube}) = \f{2L^2}{\chi} \left\{ \phi+\f{\sin 2\phi}{2}+\rho
\chi \cos\phi -2\rho \sin \phi \right\} \ ,
\eqx
where $\rho\equiv a/L$ and $\sin \phi=i\chi\, T_{tube}/L$ is
the new variational parameter.

In the strong coupling limit ($\al' =1/\sqrt{2g^2_{YM}N} \!\to\! 0$)
the parameter $\phi$ is dynamically determined from the saddle point 
equation: 
\eq
\label{e.sp}
0=\f{\partial Area(\phi)}{\partial \phi}=
\cos\phi(\cos\phi-\rho)-\f{\rho \chi}{2}\sin\phi
\eqx
It is easy to realize that for large enough energy, the last term dominates and 
thus  $\phi \sim \pm n\pi$. Inserting this solution into the area
(\ref{e.tubeminkow}) we find 
\eq
\label{e.aphi}
Area(\phi)=-\f{2L^2}{\chi} n\pi+2aL (-1)^n
\eqx
where we retain the physical solutions with $n$ positive integer. We
thus find a set of solutions very similar to the inelastic factor
obtained in the previous section. The modification due to the front-back
contribution $2aL$ is negligible in the Fourier transformed amplitude
for momentum transfer $\sqrt{q^2} \gg a/R_0^2$. Also this term is
probably more dependent on the treatment of the front-back parts of
the tube in our approximation.

\FIGURE{\epsfig{file=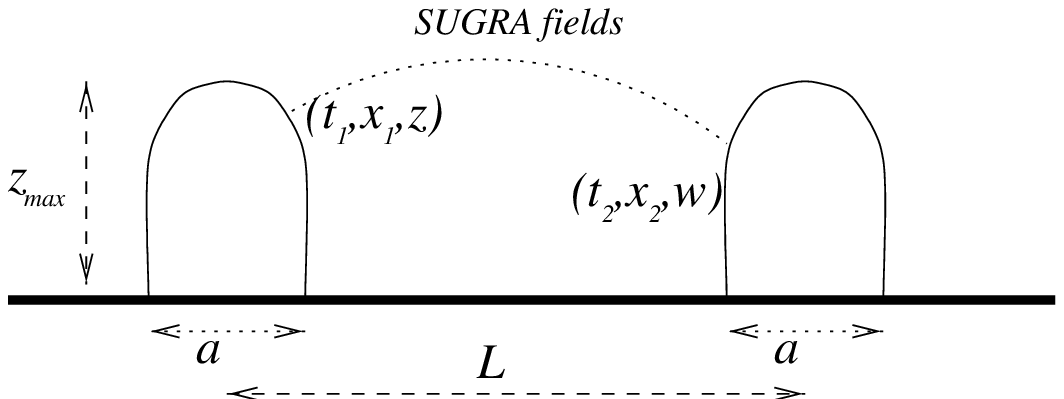,angle=0,width=13cm}%
 \caption{Dipole-dipole scattering at large impact parameter.}%
	\label{5}}

Concerning the non-confining metrics $AdS_5 \times S_5,$ the minimal area 
solution with the corresponding  boundary conditions  
is difficult to find in analytic form, necessary for the continuation to 
Minkowski space. However, there exists a generic   and intringuing 
feature: the 
existence of a  geometrical  
transition 
between small and  large impact parameter, corresponding to the 
realization 
of the minimal surface by two 
disconnected ones, see Fig.\ref{5}, where each of them reproduces the 
known 
solution used for the  calculation of the interquark potential \cite{12}.

Taking advantage of this unique configuration, valid at large enough 
impact 
parameter distance, it is possible \cite{us1} to compute the elastic 
amplitude 
{\it via} the supergravity approximation of the AdS string theory at small 
curvature. The amplitude is dominated by the exchange contribution of all 
zero-mass excitations of the appropriate supergravity theory, namely,  
Kaluza-Klein scalars, dilaton, antisymmetric tensor and graviton. All in 
all, 
the graviton dominates at large energy as $A(s,L) \propto s \times L^{-6}$ 
in 
the amplitude, but the region of validity of the supergravity 
approximation 
requires a  condition $L\gg s^{2/7} $ or $A(s,L) \ll s^{-5/7}$ 
which 
lies significantly below the absolute unitarity limit $A(s,L) < {\cal 
O}(1).$
As expected the behaviour at large $L$ is power-like and, for fixed $s$ is 
found 
dominated by the KK scalar tail in $s \ L^{-2}.$

\subsection{Dipole-dipole   inelastic scattering}
\label{rub}
The application of AdS/CFT correspondence for the two previous exemples 
is 
not 
so easy, even if partial results are encouraging. For ``quark'' elastic 
scattering, an infra-red time-like cut-off is to be introduced due to 
the 
colour 
charges of the quarks which implies a regularization scheme and a 
complication 
of the geometrical aspects.  For dipole  
elastic scattering, there is no need for a cut-off but the geometry of 
the 
minimal surface is complicated. Inelastic scattering of dipoles allows 
one 
to 
circumvent both of these difficulties. Indeed, the helicoidal geometry remains  
valid 
due 
to the eikonale approximation for the ``spectator quarks''
while the ``exchanged quarks'' define a trajectory drawn on the 
helicoid, see 
Fig.\ref{6}. This trajectory plays the r\^ole of a dynamical time-like  
cut-off which takes part in the 
minimization 
procedure.

\FIGURE{\epsfig{file=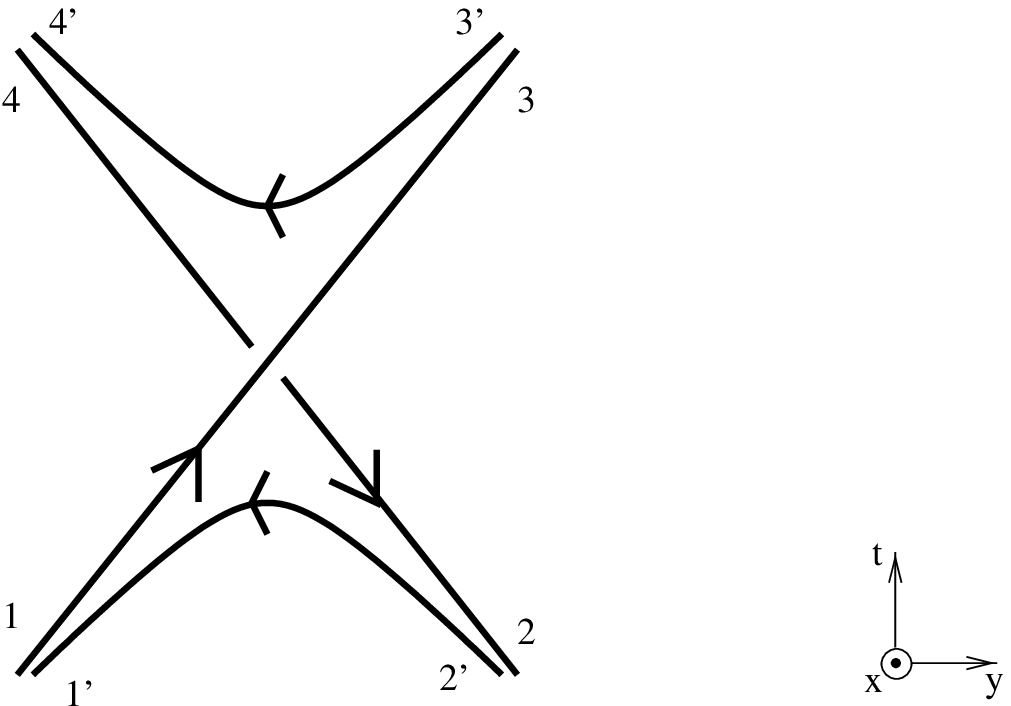,angle=0,width=10cm}%
 \caption{Wilson lines for inelastic dipole scattering.}%
	\label{6}}

Following the approach of \cite {us4} let us consider a meson-meson 
scattering process
\eq
\label{e.scat}
(11')+(22') \lra (33')+(44'),
\eqx
where the continuous lines 1-3, 2-4  correspond to spectator quark and
antiquark, 
while the dashed lines 1'-2', 3'-4' correspond to annihilated and produced
quark-antiquark pairs. The labels correspond to the initial and final
spacetime position 4-vectors that we fix for our calculation.

The spacetime picture of this process is schematically illustrated
in figure (\ref{6}), where the impact parameter axis $x$ is perpendicular 
to the
longitudinal $t-y$ plane. Note that the impact parameter is defined w.r.t. 
the 
spectator quark asymptotic trajectories.

The amplitude 
corresponding to the scattering process (\ref{e.scat}) can be
schematically written as 
\eq
\label{e.ampgen}
\cor{ <\!\!out|
S_F(3',4'|\AA)\, S_F(1,3|\AA)\, S_F(4,2|\AA)\, S_F(2',1'|\AA) 
|in\!\!> }_\AA
\eqx
where $<\!\!out|$ and $|in\!\!>$ are wavefunctions for the outgoing
and incoming mesons (up to modifications due to LSZ reduction
formulae). In formula (\ref{e.ampgen}),
$S_F(X,Y|\AA)$ denotes the
full quark propagator between spacetime points $X$ and $Y$ 
in a given background gauge field configuration $\AA$,
while the correlation function $\cor{\ldots}_\AA$ stands for averaging
over these configurations. 

Let us first perform the calculations for the above
scattering amplitude rotated into Euclidean space. 
In impact parameter space we use 
the worldline expression for the (Euclidean) fermion propagator in a
background gauge field $\AA = A^C_\mu(X^\mu)$ as a path
integral over classical trajectories \cite{15}:
\eq
\label{e.s}
S_F(X,Y|\AA)= \int_0^\infty dT e^{-m T} \int \DD X^\mu(\tau) 
\dl(\xd^2\!-\!1)\, I[X^\mu(\tau)]\, P e^{i\int A_\mu(X(\tau))\cdot \xd^\mu 
d\tau} 
\eqx
Here the path integral is over trajectories $X^\mu(\tau)$ joining $X$
and $Y$, parametrized by $\tau\in (0,T)$. Because of the delta
function, $T$ is also the total length of the trajectory. 
The quark mass dependence appears in the first exponential.
The colour and gauge field dependence is encoded in the
(open) Wilson line along the trajectory $P e^{i\int A_\mu(X(\tau))\,
\xd^\mu d\tau}$, while the spin 1/2 character 
of the quark is responsible for the appearance of the spin factor:
\eq
\label{e.sfdef}
I[X^\mu(\tau)]=P \prod \f{1\!+\!\xd^\mu \gm_\mu}{2}
= \lim_{N\to \infty} 
\f{1\!+\!\xd^\mu(T) \gm_\mu}{2}  \ldots  \f{1\!+\!\xd^\mu(\f{2}{N}T)
\gm_\mu}{2} \f{1\!+\!\xd^\mu(\f{1}{N}T) \gm_\mu}{2}
\eqx
where the second equality gives a suitably regularized definition of the 
infinite 
product
along the trajectory $X^\mu(\tau)$.
Note that each of the $N$ factors in this expression is
a projector due to the fact that $\xd^2=1$.
This spin factor was first formulated for D=3  and later
for arbitrary D \cite{15}. 
In practice it was computed explicitly in D=2 and D=3, but not in
general for $D>3$. We computed it \cite{us4} for the
configuration of figure  (\ref{6}), {\it i.e.} in a $D=3$ submanifold in 
$D=4$ spacetime. 

Let us  comment two important steps \cite{us4} of the calculation  of 
(\ref{e.ampgen}).
   
{\bf i)} Since the initial and final mesons are colour singlets, the four
Wilson lines   close to
form a single Wilson loop, and the gauge
averaging factorizes out of the expression:
\eq
\cor{\tr P e^{i\int_C \vec A \cdot d \vec X}}_\AA
\eqx
where the contour $C$ follows the quark trajectories  $1\to 3' \to
4' \to 2' \to 1'$ (following the contours sketched  on Fig.\ref{6}). Hence, 
adopting
 the ``world-line'' path integral scheme of Feynman \cite{15}, 
one may write 
the 
inelastic amplitude in terms of a Wilson loop {\it vev}:
\eq
\int \DD\tau\, \cor{W(1\!\to\! 3'\! \to \!4'\! \to \!2' \!\to \!1')}_
{\AA,\tau} 
\ e^{ -2m {\cal L}(\tau)} \ ,
\label{tau}
\eqx
where $\tau$ parametrizes the boundary trajectories and ${\cal L}$ is 
their 
total length. Using the AdS-CFT correspondence in the same framework as 
previously,  
one may formally integrate   over  the gauge degrees of freedom and 
write 
\eq
\cor{\tr P e^{i\int_C \vec A \cdot d \vec X}}_\AA \equiv \cor{W(1\!\!\to\!\! 
3'\!\! \to \!\!4'\! \!\to\! \!2' \!\!\to 
\!\!1')}_{\AA,\tau}\!\! =\! 
e^{-\!\f{Area(\tau)}{2\pi\alpha'} } \times Fluct(\tau).
\eqx
Note that the remaining minimization of (\ref{tau}) in $\tau$ runs now on both  
the 
area  and its 
boundary.

{\bf ii)} The spin factor matrices multiply 
\eq
I[1\!\to\! 3]_{\al_1\al_3}\, I[4\!\to\! 2]_{\al_4\al_2}\,
I[2'\!\to\! 1']_{\al_{2'}\al_{1'}}\, I[3'\!\to\! 4']_{\al_{3'}\al_{4'}} 
\eqx
and are contracted with the  initial and final spinor
wavefunctions like $u_{\al_1}(p_1)\vb_{\al_{1'}}(p_1),$ corresponding to a 
simple approximation for the wave-functions of the external mesons as
mentioned in the introduction.   
After non-trivial simplifications due to the 3-dimensional dimension of 
the embedded trajectories, 
one finds
\eq
I[\xd] =
\f{1\!+\!\xd^\mu(T) \gm^\mu}{2} \f{1\!+\!\xd^\mu(0) \gm^\mu}{2} \cdot
\left(\f{1\!+\!\xd(T)\cdot\xd(0)}{2}\right)^{-1}\Rightarrow \f 1s \ ,
\eqx
once contracted with the initial and final spinors.

Let us  focus on the 
configuration 
of Wilson lines of Fig.\ref{6} in the context  of a confining theory. As 
previously noted, the main contribution to 
the 
minimal area is from the  metrics in the bulk near $z_0$ which is nearly
flat. Hence, near $z_0,$ the relevant minimal area can be  drawn on a
classical helicoid. However, by contrast with the previous cases, the 
natural cut-off is provided by the exchanged quark trajectory, which  is 
self-consistently fixed by the minimization procedure. 
The solution of the amplitude boils down to an Euler-Lagrange  
minimization 
over $\tau,$ namely
\eq
A_{\cal R}(s,L^2)\propto\f{1}{s}\lim_{\alp\to 0}\int \DD\tau\,  
e^{-\f{1}{2\pi\alpha'} Area(\tau)}
 e^{-2m {\cal L}(\tau)} \times Fluct. \ ,
\label{e.pathint}
\eqx
where $Area(\tau)$ is the section of an helicoid bounded by the quark 
trajectories  having total length ${\cal L}(\tau).$

It can be easily shown that the Euler-Lagrange equations admit a solution which 
mimimizes both the area and the boundary length, namely
\eq
\label{e.el}
\f{\partial (-\f{1}{2\pi\alpha'}Area\!-\!2m 
Length)}{\partial \tau} = 0\Rightarrow  \sqrt{1+\left(\f 
{\theta\tau}{L}\right)^2}=0
\ .
\eqx 
The solution is a  constant ($\f{\partial \tau(\sg)}{\partial \sg}=0$) and 
complex trajectory 
\eq
\tau(\sigma)_{min}\equiv \Tau=\pm i L/\theta\ .
\label{constant}
\eqx
 Here the complex
value has to be understood in the sense of applying the steepest descent
method to the path integral (\ref{e.s}), and deforming the
integration contours into the complex plane.

Substituting the classical solution $p\tau(\sg)=-i$ into
(\ref{e.s}) gives a non vanishing contribution from the
logarithm:
\eq
\label{e.spa}
e^{-\f{1}{2\pi\alef} Area(-iL/\theta)} = e^{-\f{i L^2}{4\alef \th}} \lra 
e^{-\f{ L^2}{4\alef \chi}} 
\eqx
after analytical continuation to
Minkowski space. 

Performing the Fourier transform, the 
resulting amplitude reads:
\eq
A_{\cal R}(s,q^2)=\int d\vec l\ e^{i\vec q\cdot \vec l}\ e^{-\f{ 
L^2}{4\alef 
\chi}}\propto  s^{-\alef q^2}\ ,
\label{reggeon}
\eqx
corresponding to a linear Regge trajectory with intercept $0$ and 
slope $\alef$ related to the quark potential calculated within the same 
AdS/CFT 
framework. 

\section{Beyond the classical approximation: Fluctuations }

\FIGURE[hb]{\epsfig{file=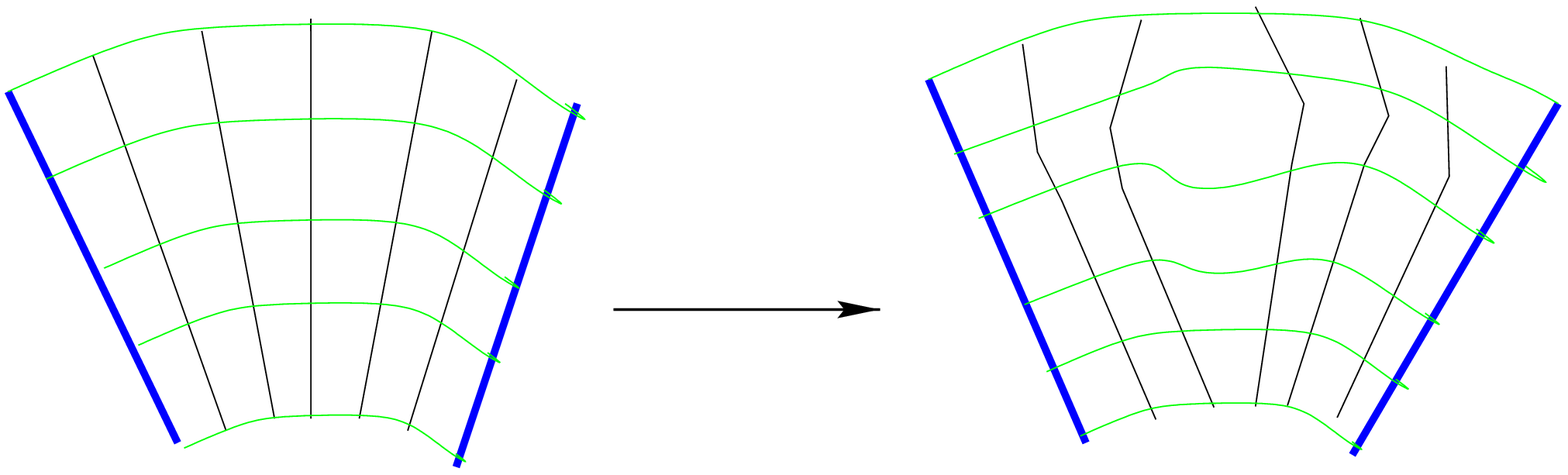,angle=0,width=12cm}%
 \caption{Fluctuations around the minimal helicoid.}%
	\label{7}}

Up to now, we restricted ourselves to  a classical approximation based on the  
evaluation of minimal surfaces solutions for the various Wilson loops involved 
in the preceeding calculations. It is interesting to note \cite {us3} that a 
further step can be done by evaluating the contribution of quadratic 
fluctuations of the
string worldsheet around the minimal surfaces in the case where these surfaces 
are embedded in helicoids, as discussed for the confining backgrounds.
The semi classical correction   comes from the fluctuations near the 
minimal surface sketched 
in Fig.\ref{7}. The main outcome is that this semi classical correction can be 
computed and   is intimately related to the well-known ``universal'' 
L\"uscher 
term contribution to the  interquark potential \cite{16}.

The fluctuation determinant for the case of a helicoid bounded by two
helices with $\tau=\pm \Tau$ has already been calculated \cite{us4,us3}. Let us
briefly recall the basic steps. First one reparametrizes the helicoid  by 
replacing 
the variable $\tau$ in (\ref{helico}) by
\eq
\rho=\f{L}{\th}\ \log\left(\f{\th\tau}{L} + \sqrt{1+\f{\th^2\tau^2}{L^2}}\right) 
\ .
\eqx
In the variables $\rho$, $\sg$ the induced metric on the helicoid has a 
conformal factor {\it i.e.} 
\eq
g_{ab}=(\cosh^2 \th\rho/L)\ \ \dl_{ab} \ .
\eqx
Therefore,
since string theory in the AdS background is expected to be {\em critical} 
(conformal 
invariant), we may
perform the 
calculation getting rid of the conformal factor, {\it i.e.} for the conformally 
equivalent  flat metric $g_{ab}=\dl_{ab}$. This 
reduces to a
calculation of the fluctuation determinant for a rectangle of size
$a\times b$ where
\eq
a = L\ ;\ 
b = \f{2L}{\th} \log \left( \th\Tau/L+\sqrt{1+\th^2 \Tau^2/L^2} \right)\ .
\eqx 
 Furthermore, we assume that the quadratic bosonic fluctuations are
governed by the Polyakov action, as is indeed the case for string
theories on AdS backgrounds. For high energies (after continuation to
Minkowski space at large $a/b ={\cal O}(\log s) \gg 1$) one obtains  
\eq
\label{e.fluct}
Fluct.(\tau(\sg)\equiv\Tau)\to
\exp \left( \f{n_\perp \cdot \pi}{24} \cdot
\f{a}{b} \right)
\ ,
\eqx
where $n_\perp$ is the number of zero modes in the transverse-to-the-branes 
directions.
The result is just equivalent to the  L\"uscher term in the potential
(c.f. Ref.\cite {16}) except that the number of zero modes $n_\perp = D-2$ can 
be larger than 
the  usual value (2) corresponding to  flat $4D$ space.

Let us consider the resulting amplitudes after account taken of  the fluctuation 
contributions:

For elastic dipole-dipole scattering, see the discussion in subsection 
\ref{dub}, one considers \cite{us4,us3} the  analytic continuation of the 
minimal area  when $\th \to \chi/\pi.$ Retaining here also the dominant term in 
the $b \to  \f{4i\pi L}{\chi} $ for large 
$\chi \sim \log s,$ one obtains
the fluctuation-corrected ``Pomeron'' amplitude
\eq
A_{\cal P}(s,t) \propto s^{\al_{\cal P}(t)}=s^{1+\f{n_\perp}{96}+\f{\alef}{4} t}
\ .
\eqx

For two-body inelastic scattering, see the discussion in subsection \ref{rub},  
one has to implement the minimal condition (\ref{constant}) namely  $b\equiv  
\f{i\pi L}{\chi}.$ Hence, 
the  fluctuation-corrected ``Reggeon'' trajectory (cf. 
\ref{reggeon})
reads\footnote{
Possible logarithmic
prefactors,  which are not under control at this stage of our 
approach, are not determined.}:
\eq
\label{e.lintraj}
A_{\cal R}(s,t) \propto s^{\al_{\cal R}(t)}\ =s^{\f{n_\perp}{24}+{\alef}t}
\ .
\eqx

Let us comment these results. 
The first observation is that in both cases, the slope is determined
by minimal surface solutions through the logarithmic contribution in
the helicoid area. The factor four  in the slope
comes from the specific saddle point path integral over the exchanged quark
trajectories (for Reggeon exchange). It is interesting to note that
this theoretical feature is in qualitative agreement with the phenomenology of
soft scattering. Indeed once we fix the $\alef$ from the
phenomenological value of the static $q\qb$ potential ($\alef\sim 0.9\,
GeV^{-2}$) 
we get for the slopes $\al_R=\alef\sim 0.9\, GeV^{-2}$ and
$\al_P=\alef/4\sim 0.23\, GeV^{-2}$ in good agreement with the phenomenological
slopes.

The second feature is the relation between the Pomeron and Reggeon
intercepts. At the classical level of our approach these are
respectively 1 and 0. Note that this classical piece is in
agreement with what is obtained  from simple exchanges of two
gluons and quark-antiquark pair, respectively, in the $t$ channel. The 
fluctuation (quantum) contributions to the Reggeon and Pomeron are also
related by the factor four. 
  
Adding both classical and fluctuation contributions gives an estimate
which is in qualitative agreement with the observed intercepts.  
Indeed, when calculating the fluctuations around 
a minimal surface near the horizon in the
BH backgrounds there could be  $n_\perp=7,8$ massless bosonic modes 
\cite{so99}.  For $n_\perp=7,8$ one gets $1.073-1.083$ for the Pomeron and 
$0.3-0.33$
for the Reggeon. 
This result is in agreement with the observed intercept for the ``Pomeron'' and 
somewhat below the intercepts of around $0.5$ observed for the
dominant Reggeon trajectories. 

An interesting feature of the results  is the
key role of the logarithmic term in the formulae (cf. (\ref{e.ahelic}) for the 
area of the truncated helicoids. Besides the main feature being that it leads, 
through its analytical structure, to  Reggeization, it
also gives rise to the possibility of 
additional contributions from crossing  different Riemann sheets
($log \to log+2\pi i k$)
in the course of performing analytical continuation from Euclidean to
Minkowski space . 

For instance in the ``Reggeon'' case,
the amplitude in impact parameter space (\ref{reggeon})  picks
up new multiplicative factors:
\eq
\label{e.rp}
e^{-\f{L^2}{4\alef \log s}} \cdot e^{-k\f{L^2}{\alef \log s}} \ .
\eqx
This can be interpreted (for $k>0$) as $k$-Pomeron exchange 
corrections to a single Reggeon exchange. Indeed the slope of the
trajectory obtained from Fourier transform of formula (\ref{e.rp}) is
 the one expected from such contributions\footnote{However, the semi-classical 
correction to the intercepts seems to be more delicate, and needs further 
study.}. 

\section{Conclusion: Reggeization and confinement}

The interesting output of the application 
of 
AdS/CFT correspondence to high energy amplitudes at strong coupling is 
to 
emphasize the relation between Reggeization and confinement, using the 
description of two-body scattering amplitudes in the dual string theory. Lattice 
calculations, which is the only presently known way to evaluate 
directly QCD observables at strong coupling, are not able to compute 
high-energy amplitudes.

When comparing  AdS$_5$  duality - which corresponds to a conformal, 
non-confining gauge theory - with   AdS$_{BH}$  duality, which leads 
to 
reggeization, the difference ultimately comes from the different metrics 
in 
the bulk and thus from the different geometrical features of the minimal 
surfaces for the same boudary 
conditions. In particular, taking into account  their different geometry, see 
e.g.  
Fig.\ref{8}, one expects after analytic continuation and in the 
large 
energy ($\chi\to \infty$) limit:
\eq
Area^{AdS}_{min}\sim 
\lim_{\chi\to \infty} \f{L}{L/\chi}\ ; \ Area^{BH}_{min}\sim 
\lim_{\chi\to 
\infty} 
L\times\f{L}{\chi}\ .
\eqx

\FIGURE[ht]{\epsfig{file=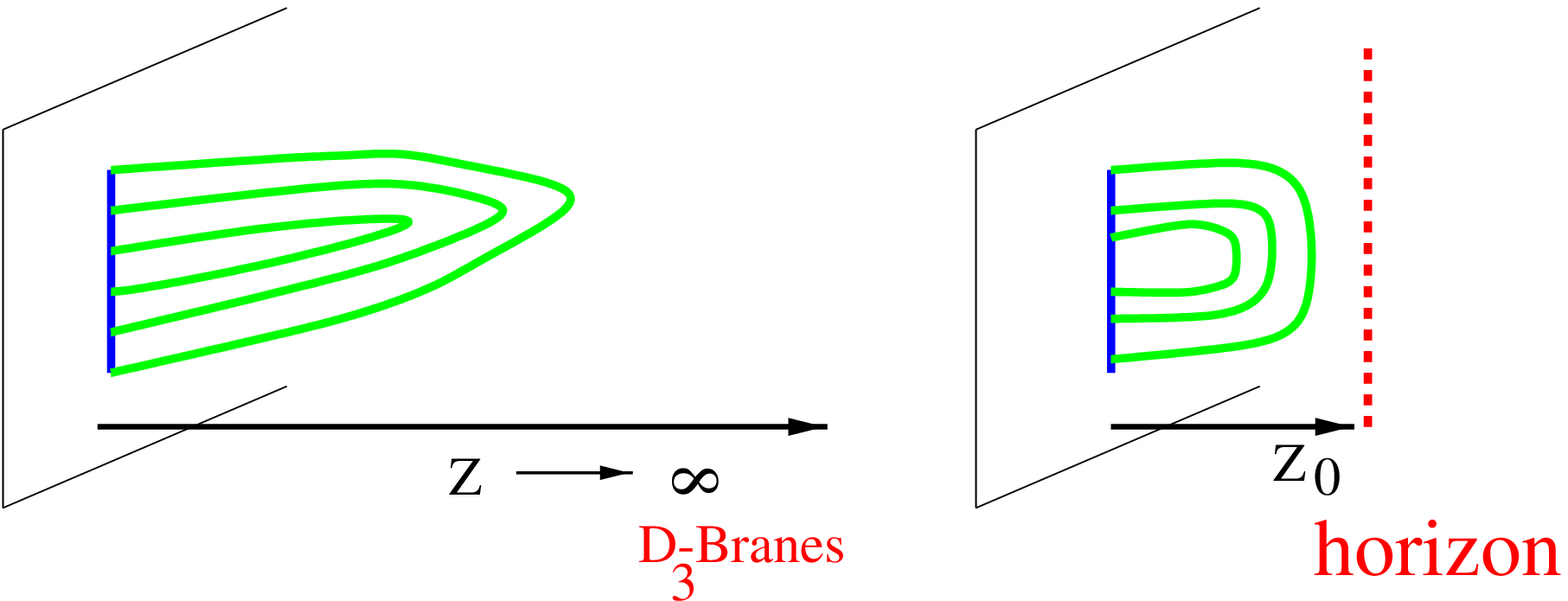,angle=0,width=12cm}%
 \caption{Comparison of $AdS_5$ and $AdS_{BH}$ minimal contours at high energy.}
 \label{8}} 
 
The AdS$_5$  case leads to a $L$-invariant value, as it is scale  invariant,  
and, 
after 
Fourier transformation, to a high-energy amplitude with a $q^2$ 
independent 
energy exponent (or flat Regge trajectory), see formula (\ref{e.lcft}). On the 
other hand, the 
AdS$_{BH}$ 
case  leads to a linear Regge trajectory 
after 
Fourier transformation. 
For the AdS$_{BH}$ case this rough expectation can be verified by
an explicit calculation. Hence confinement appears as an essential 
ingredient 
for the reggeized structure of two-body high-energy amplitudes. We expect this 
result 
 not to be dependent on the precise geometrical 
AdS$_{BH}$ setting and thus to indicate a quite general property of 
confining 
theories.

As a conclusion, let us summarize our results:
\begin{itemize}
\item The AdS/CFT correspondence can be used to give a geometrical formulation 
of two-body scattering amplitudes in the gravitational dual of  gauge field 
theories.
\item The consideration of  quark-quark scattering in physical Minkowski space 
shows that the main geometrical features of two-body scattering amplitudes are 
related to a (generalized to AdS metrics)  helicoidal structure of minimal 
surfaces in Euclidian space {\it via} analytical continuation.
\item In the case of non conformal theories, such as the $AdS_{BH}$ case, the 
metrics is approximately flat near the horizon corresponding to the confinement 
scale. The (flat space) helicoidal solutions lead to  amplitudes with linear 
Regge slopes.
\item While quark-quark scattering is entailed by a cut-off dependence, 
colourless dipole scattering give rise to cut-off free amplitudes. In particular 
the two-body  dipole-dipole scattering with quark exchange leads to unambiguous 
results with well-defined regge behaviour.
\item Regge trajectories come out  linear, 
with 
slopes and intercepts related to the quark potential. They  include a 
semi-classical correction due to the fluctuation around the minmal surfaces 
which are similar to  a  
L\" uscher 
term, but in a 10-dimensional string framework.
\item The  Pomeron (elastic case) intercept is  $1+\epsilon$ where  $\epsilon$ 
is related to a L\" usher term. There exists a factor four between the Reggeon 
(inelastic case) and Pomeron Regge slopes in agreement with ``soft scattering'' 
phenomenology.

\end{itemize}

In conclusion,  the AdS/CFT framework give new insights on the  35-years-old 
puzzle of high-energy amplitudes at strong gauge coupling.

As a short outlook let us list some interesting problems for future work:
 
\begin{itemize}
\item {\it High-energy phenomenology:} Many aspects, like 
the 
Flavor/Spin dependences,  remain  to be studied.  
\item {\it Approximations:} The dual gauge theory is not 
specified, and the exact minimal surface in the bulk metrics to be 
determined. 
\item {\it Dual of QCD?} In the present framework, the 
confining  scale $R_0$ has no relation with  $\Lambda_{QCD}.$
\item {\it Unitarity:}  A more complete investigation  
requires the study of multi-leg amplitudes.
\item {\it Deeper general problems:} The formulation of 
string 
theory in AdS backgrounds and last but not least, a proof of the AdS/CFT
conjecture.

\end{itemize}

\acknowledgments

I warmly thanks Romuald Janik  with whom the   approach of  high-energy 
amplitudes described in this review has been done in tight collaboration. I 
thank Otto Nachtmann and the organizers of the Heidelberg meeting for the 
stimulating and fruitful atmosphere.

\end{document}